# Fast transient charge trapping in salt-aided CVD synthesized monolayer MoS₂ field-effect transistor

Sameer Kumar Mallik,[1,2] Sandhyarani Sahoo,[1,2] Mousam Charan Sahu,[1,2] Sanjeev K Gupta,[3] Saroj Prasad Dash,[4] Rajeev Ahuja,[5,6] Satyaprakash Sahoo[1,2*]

[1]Laboratory for Low Dimensional Materials, Institute of Physics, Bhubaneswar-751005, India

[2]Homi Bhaba National Institute, Mumbai-400094, India

[3]Computational Materials and Nanoscience Group, Department of Physics and Electronics, St. Xavier's College, Ahmedabad 380009, India

[4]Department of Microtechnology and Nanoscience, Chalmers University of Technology, SE-41296, Göteborg, Sweden

[5]Condensed Matter Theory Group, Department of Physics and Astronomy, Box 516, Uppsala University, S-75120 Uppsala, Sweden

[6]Applied Materials Physics, Department of Materials and Engineering, Royal Institute of Technology (KTH), S-100 44 Stockholm, Sweden

## Abstract

Atomically thin semiconductors have versatile future applications in the information and communication technologies for the ultimate miniaturization of electronic components. In particular, the ongoing research demands not only a large-scale synthesis of pristine quality monolayer MoS₂ but also advanced nanofabrication and characterization methods for investigation of intrinsic device performances. Here, we conduct a meticulous investigation of the fast transient charge trapping mechanisms in field-effect transistors (FETs) of high-quality CVD MoS₂ monolayers grown by a salt-driven method. To unfold the intrinsic transistor behavior, an amplitude sweep pulse I~V methodology is adapted with varying pulse widths. A significant increase in the field-effect mobility up to ~100% is achieved along with a hysteresis-free transfer characteristic by applying the shortest pulse. Moreover, to correlate these results, a single pulse time-domain drain current analysis is carried out to unleash the fast and slow transient charge trapping phenomena. Furthermore, rigorous density functional theory (DFT) calculations are implemented to inspect the effects of the Schottky barrier and metal-induced gap states between drain/source electrode and MoS₂ for the superior carrier transport. Our findings on the controllable transient charge trapping mechanisms for estimation of intrinsic field-effect mobility and hysteresis-free transfer characteristic in salt-assisted CVD-grown MoS₂ FETs will be beneficial for future device applications in complex memory, logic, and sensor systems.

* Corresponding Author: sahoo@iopb.res.in

# Introduction

Atomically thin two dimensional (2D) layered materials,[1] monolayer $MoS_2$ in particular, has opened a new window in miniaturizing electronic devices owing to its excellent optical, thermal, and electrical properties.[2–4] Having a direct bandgap of 1.8 eV,[5] large exciton binding energy, and strong spin-valley coupling, it offers a wide range of potential applications in the field of optoelectronics,[2] spintronics,[6] valleytronics, and quantum optics.[7] Researchers have followed numerous methods for the qualitative synthesis of monolayer $MoS_2$ among which chemical vapor deposition (CVD) technique is the most successful and reliable method when it comes to large scale (of about several tens of microns), good crystalline quality, and thickness dependent growth.[8,9,18,10–17] In the conventional CVD process, it has been observed that the growth conditions, size, and shape of the $MoS_2$ domains are greatly affected by numerous parameters such as growth temperature, growth time, pressure, flow rate, amount of loaded precursors, etc. Nevertheless, the relatively low vapor pressure of metal precursor leads shortening of grain size and also makes complicate to control the aforementioned parameters during growth time. To suppress the nucleation and promote the lateral growth dynamics, the assistance of salt (e.g., NaCl, KCl) to the metal precursor plays a pivotal role in enhancing the growth rate even at atmospheric pressure.[14–16] However, the addition of NaCl not only triggers the relative vapor pressure of the reactants but also lowers the growth temperature as compared to the conventional CVD process.[12,13] Recently, Zhou *et al*[11] have reported NaCl aided controlled synthesis of several transition metal chalcogenides (TMDs). The kinetics of the reaction mechanism lies in the fact that with the presence of NaCl during the reaction, $MoO_3$ forms an intermediate state as oxychloride e.g. $MoO_xCl_y$, which drastically reduces the melting point as compared to pure $MoO_3$. For the past few years, salt-driven synthesis of layered 2D materials such as $MoS_2$ and $WS_2$ has proven to be a viable method due to their critical control over thermodynamic stability during vapor phase reactions.

Coming to the widespread applications, $MoS_2$ based field-effect transistors (FETs) have paramount importance in future electronic applications such as communication, data storage, memory, optoelectronics, quantum optics, etc. owing to their high electron-hole mobilities, high ON/OFF current ratios, low operation power, good photoresponsivity, high sensitivity for molecular detection, etc.[19–23] Indeed, a typical room temperature mobility range of about 0.8-18 $cm^2V^{-1}s^{-1}$ is very common in an unencapsulated monolayer $MoS_2$ with current ON/OFF ratios in order of $10^5$-$10^7$.[8–10,18] However, an unusual hysteresis in the transfer curves of $MoS_2$ based FETs plays a critical factor in regulating the electrical performance of the devices such as threshold voltage instability, low carrier mobility, drain current reduction, etc.[24–26] In literature, several remarks on the origin of hysteresis have been discussed to avoid or minimize such unwanted behaviors in $MoS_2$ FETs.[24,27–29] The carrier trapping mechanism, which is considered to be a major factor for hysteretic behavior, is being studied widely to control the broadening of the hysteresis.[30–32] However, the majority of the previous works on hysteresis studies cover a few layers of exfoliated $MoS_2$, and a very limited study on CVD-grown monolayer samples is recently reported.[33] Moreover, a comprehensible understanding of the localized interface trap states using the pulsed I~V technique is still in its infancy. Although the newly developed salt-assisted CVD-growth has emerged as a prevalent method for producing high-quality monolayer $MoS_2$, the transient charge trapping mechanisms and hysteresis-free transfer characteristics have not been reported so far. Furthermore, the estimation of intrinsic field-effect mobility from the hysteresis-free transfer characteristic in salt-driven CVD-grown $MoS_2$ FETs is still lacking.

In this report, we demonstrate the time-domain drain current responses to minimize the fast transient charge trapping effects on a NaCl-assisted CVD synthesized high-quality monolayer $MoS_2$ FETs. DFT study is performed to explore the interfacial properties between drain/source electrode (Ag) and monolayer $MoS_2$. The room temperature hysteresis studies using DC and pulsed I~V techniques are carried out in both ambient and vacuum conditions to inspect various trapping mechanisms to extract intrinsic FET parameters. An increase in the field-effect mobility up to 100% is achieved in short pulse amplitude sweep measurement than DC measurements. To further explore, single-pulse transient measurements are carried out to get insight into the fast trapping phenomena. Our studies reinvigorate the origin of hysteresis in salt-aided monolayer $MoS_2$ samples; further understanding these significant trapping mechanisms is of prime importance for the next-generation device applications.

## Results and Discussion

Single-layer $MoS_2$ is prepared by NaCl-driven CVD technique in a single zone tube furnace directly on Si/$SiO_2$ substrate (see Methods section for more details). NaCl is introduced to increase the vapor pressure of the metal precursor ($MoO_3$) and promote the lateral growth dimension.[11] The hexagonal structures are favored with the growth parameters and have a much higher yield than the regular triangular domains. Figure 1 (a) shows an optical micrograph of a defect-free hexagonal domain single-crystal monolayer $MoS_2$. The layer number is confirmed by the color contrast microscopic images followed by the Raman characterization. Figure 1 (b) shows the in-plane and out of plane Raman modes ($E^1_{2g}$ and $A_{1g}$, respectively) for monolayer and multilayer CVD-grown $MoS_2$. The frequency difference ($\Delta$) between the two modes can be used as an authentic identification for monolayer $MoS_2$.[34]

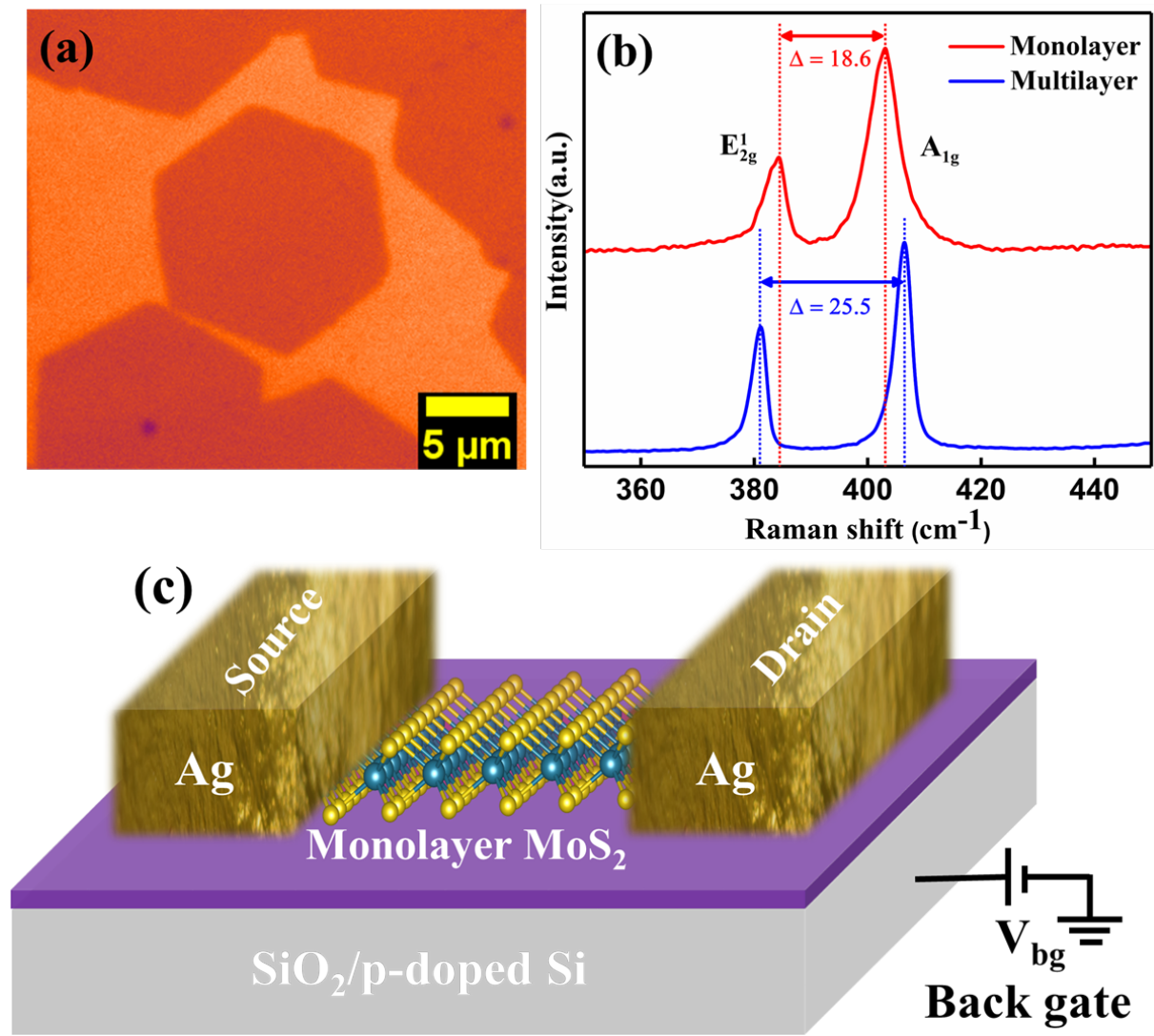

**Figure 1. (a)** *Optical microscopic image of NaCl aided CVD-grown monolayer $MoS_2$ in a hexagonal domain on $Si/SiO_2$ substrate.* **(b)** *Raman spectra with 514.5nm excitation wavelength showing two predominant Raman modes ($E^1_{2g}$ and $A_{1g}$) of monolayer and multilayer $MoS_2$. The Frequency difference ($\Delta$) between the two modes confirms the monolayer nature of $MoS_2$.* **(c)** *Schematic of monolayer $MoS_2$ FET with $SiO_2/Si$ back gate, Ag source/drain contacts, and utilized electrical circuits.*

Carrier transport in a FET channel is generally affected by various factors such as the van der Waals (vdW) gap between metal and semiconductor, Schottky barrier, metal-induced gap states, etc.[35] In Fig. 1 (c), a back gated monolayer $MoS_2$ FET is schematically shown with necessary electrical connections so that a carrier injection from metal to semiconductor channel material can be realized pictorially. Several metals and their combinations with varying work function values have been studied extensively as an electrode to $MoS_2$ based devices.[36] However, very few research enlighten on getting low Schottky barrier contacts with $MoS_2$ by using low work function metals such as Sc and Ti.[37] However, in that context, Ag is the least explored material with a work function value of 4.46 eV[38] (a little higher than for monolayer $MoS_2$) even though having an excellent electrical and thermal conductivity. Furthermore, Ag (111) surface has a hexagonal geometry with a lattice constant (2.93Å) nearly matches with monolayer $MoS_2$ (3.18Å) makes it a suitable candidate as a metal electrode to $MoS_2$ (for more information, see Computational details). To ensure our proposition and experimental validation, a comprehensive DFT study is implemented on the $MoS_2$/Ag interface. Figure 2 (a) and (b) show the band structure and corresponding partial density of states (PDOS) of monolayer $MoS_2$ 3×3×1 supercell. The direct bandgap of 1.7eV in our case explains a small underestimation as compared to the experimental value (1.8eV) in the case of monolayer $MoS_2$. This is well known that DFT-based calculations underestimate the bandgap compared to the experiment. The Brillouin zone folding in the case of supercell results in a direct bandgap at gamma point instead of K point.[39] The carrier injection (in this case electron) from metal to $MoS_2$ channel can be comprised of two steps as shown in Fig. 2 (e), the tunneling of the electron from metal to the $MoS_2$ underneath Ag (considered as electrode region) followed by the transport of electrons from electrode region to $MoS_2$ channel (known as scattering region). The corresponding band alignments at both regions with the vdW gap and Schottky barrier height ($\Phi_{SBH}$) are shown schematically in Fig. 2 (f). Further band structure calculation of the electrode region is performed to measure $\Phi_{SBH}$ which is nothing but the energy difference between $CBM(E_c)$/$VBM(E_v)$ of $MoS_2$ for n/p-type Schottky barrier respectively and Fermi level of $MoS_2$/Ag (111) heterostructure. The equilibrium separation distance ($d_{eq}$) between the Ag and MoS$_2$ is optimized for 2.5Å by the energy minimization. Figure 2 (c) shows the projected band profile of MoS$_2$/Ag (111) heterostructure (electrode region) where the color bar represents the percentage of orbital contributions of the MoS$_2$ underneath Ag. The $\Phi_{SBH}$ is calculated to be 153 meV by the usual definition (forming an n-type Schottky contact) which is a reasonably smaller value as compared to the other high work function metals.[36,37] Moreover, the projected band structure of MoS$_2$ is seen to retain its original form compared to the band profile of pristine monolayer MoS$_2$, as shown in Fig. 2 (a). This indicates a negligible strain effect of Ag (111) layers on MoS$_2$ due to the similar crystal structures and nearly equal lattice constants. A PDOS calculation is further carried out to gain insight into the effect of hybridization between Ag and MoS$_2$ on the orbital projections of MoS$_2$ at the bandgap region. As seen from Fig. 2 (d), a small amount of overlapped Mo-4$d$ state is noticed at the Fermi level and band-gap region when comparing with PDOS of pristine MoS$_2$. This spreading of band edge Mo states may arise due to the weakening of Mo-S covalent bonds by metal adsorption.[40] However, considering low work function metals for getting metalized states with low electrical conductivity and high work function metals having relatively high electrical conductivity but with higher Schottky barriers, Ag plays a perfect balancing role having high electrical conductivity and forming low Schottky barrier with MoS$_2$.

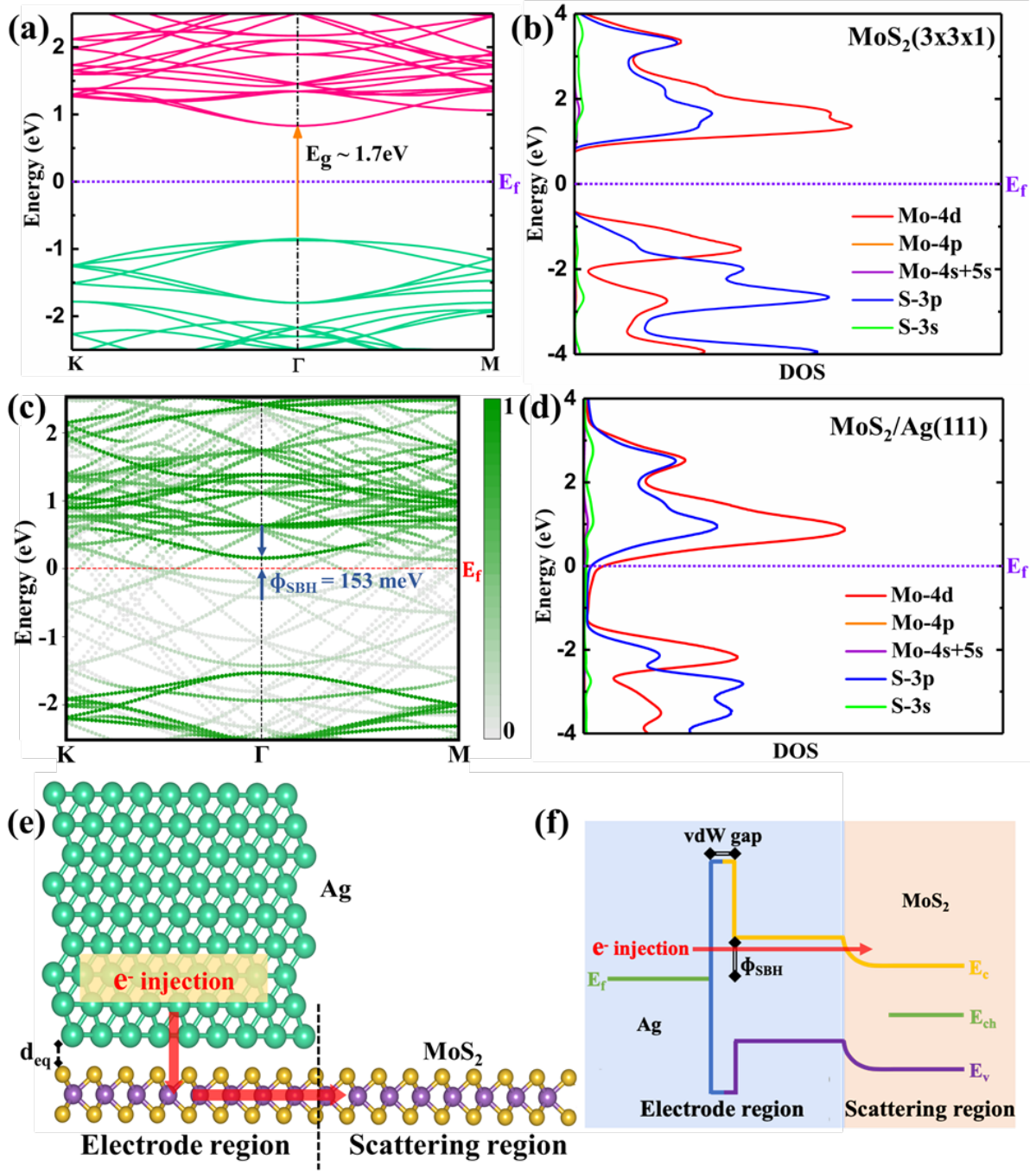

**Figure 2.** *(a) Band structure of monolayer MoS₂ supercell showing direct bandgap of 1.7eV, valence band (green solid lines), conduction band (pink solid lines), and Fermi level (E_f) (violet dotted line) are shown. (b) Corresponding PDOS showing orbital contributions at CBM and VBM. (c) Projected band profile of MoS₂/Ag (111) heterostructure showing the projection of MoS₂ bands (dotted green lines) with scalebar, The n-type Schottky barrier height (Φ_SBH) is shown in blue color. (d) Corresponding PDOS of MoS₂ showing small gap states for hybridized*

*Mo-d orbital.* ***(e)*** *Schematic of a cross-sectional view of Ag top contact to monolayer MoS₂ forming electrode and scattering region. Electron injection from Ag to MoS₂ is shown in red arrow.* ***(f)*** *Schematic illustration of band alignments near the Ag-MoS₂ interface. The red arrow shows the electron injection through the vdW gap and Schottky barrier. $E_f$ and $E_{ch}$ correspond to metal Fermi level and channel potential, respectively.*

In literature, it has been reported that Ag forms a smoother interface with monolayer MoS₂ and by using it as a metal contact, high carrier transport is achieved in a MoS₂-based transistor compared to Ti.[41] This result is consistent with our projected band structure calculations. Moreover, the lowest contact resistance is also realized by using Ag or Au as a drain/source metal contact of monolayer MoS₂ FET.[42–45] This motivated us to fabricate single-layer MoS₂ FET on the as-grown CVD sample with Ag as contact metal for drain and source electrodes by using the photolithography process. The back gate contact of the FET is facilitated by heavily doped Si substrate and 300 nm SiO₂ beneath the MoS₂ layer provides the required gate capacitance. The high magnification optical image of the as-fabricated device is shown in Fig. 3 (a). The channel length ($L$) and width ($W$) is calculated to be ~ 2.7 and 8.1μm, respectively. The inset shows a low magnification optical image of the complete FET with a scalebar of 200μm.

Nearly linear output characteristics i.e. variation of the drain current ($I_{ds}$) with drain-source voltage ($V_{ds}$) can be noticed in Fig. 3 (b), which indicates exemplary Ohmic contact behavior for the fabricated single-layer MoS₂ FET. This result is consistent with our projected band calculation. It is interesting to note that the output currents are in the range of microampere, which elucidates the high pristine quality of NaCl assisted CVD grown MoS₂ FET. The drain current is recorded for different gate voltages starting from small negative voltage (-2V) to large positive voltage (30V) shows the excellent electrostatic gate doping effect.

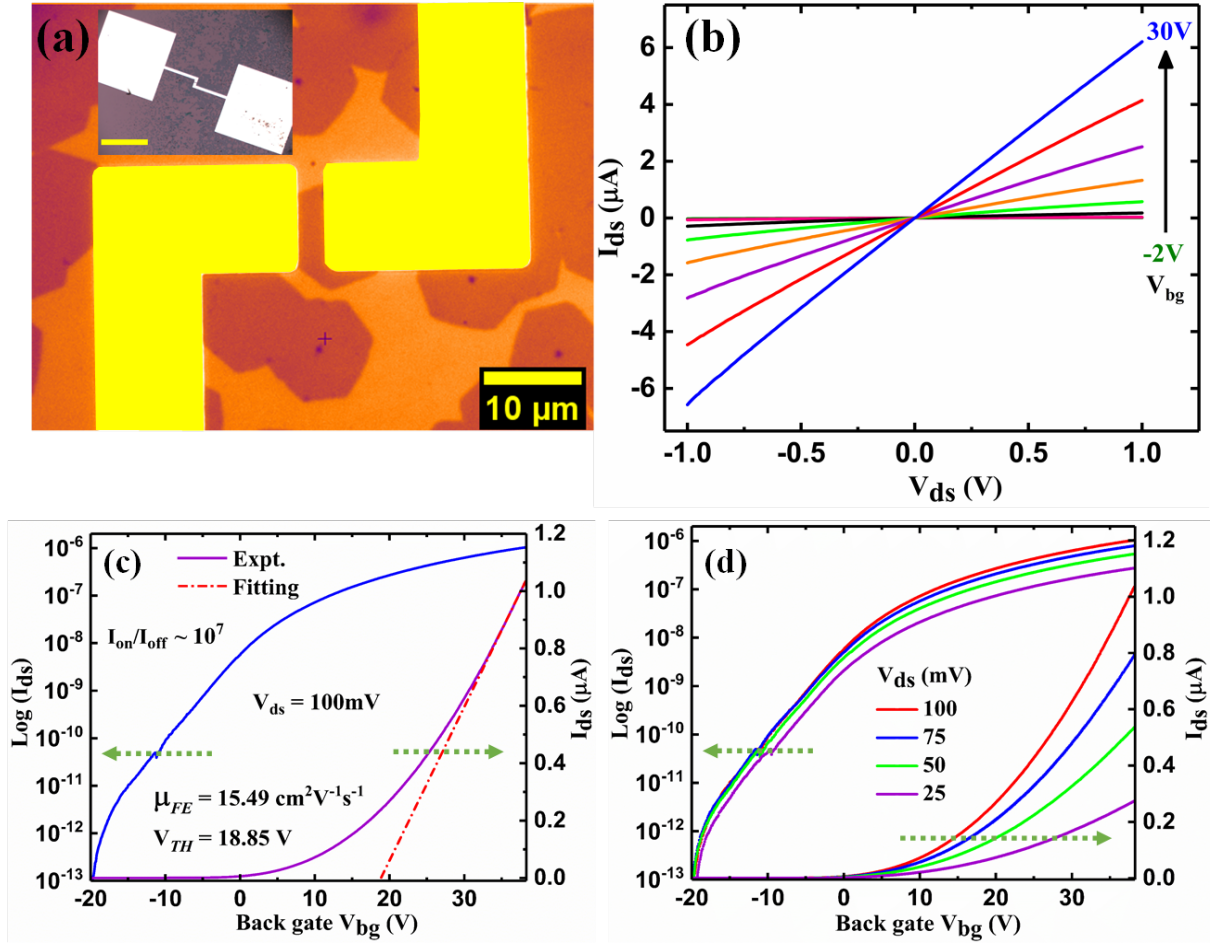

**Figure 3. (a)** *An optical microscopic image showing as-fabricated monolayer MoS₂ FET (Ag electrode is shown in false yellow color), Inset shows the low magnification image of the device with Ag contact pads (scale bar is 200μm).* **(b)** *Output characteristics i.e. drain current ($I_{ds}$) vs drain-source voltage ($V_{ds}$) with different fixed gate bias ($V_{bg}$) from -2V to 30V.* **(c)** *mobility and threshold voltage extraction from linear $I_{ds}$ transfer characteristics (right axis), corresponding semi-logarithmic drain current (left axis) showing high current ON/OFF ratio at $V_{ds} = 100mV$.* **(d)** *Room temperature transfer characteristics at different $V_{ds}$ in both linear (right axis) and semi-logarithmic (left axis) scales showing the stability of the FET.*

The electrical measurements are performed in both ambient and high vacuum (~$10^{-5}$ mbar) conditions in order to understand the effect of the surrounding environments on channel mobility. Figure 3 (c) shows transfer characteristics i.e. curve between drain-source current ($I_{ds}$) vs back gate voltage ($V_{bg}$) at a constant drain-source voltage ($V_{ds}$ = 100mV) in vacuum for a single-layer MoS$_2$ back gated FET at room temperature. The field-effect mobility of the charge carriers is extracted by using the expression[19], $\mu_{FE} = \frac{dI_{ds}}{dV_{bg}} \times [\frac{L}{WC_{ox}V_{ds}}]$ where $L$ and $W$ are the channel length and width respectively (in our case, $L$ = 2.7 μm and $W$ = 8.1 μm), $C_{ox}$ is the capacitance value of the back gate dielectric per unit area ($C_{ox} = \frac{\varepsilon_0 \varepsilon_r}{d}$; In our case, relative permittivity of SiO$_2$, $\varepsilon_r$ = 3.9, thickness of SiO$_2$ layer $d$ = 300 nm ). At a bias voltage $V_{ds}$ = 100mV, from the slope of the transfer curve, the room temperature mobility is calculated to be ~15.49 cm$^2$V$^{-1}$s$^{-1}$, which is higher or comparable to previously reported results on monolayer MoS$_2$ without surface passivation by high-k dielectrics.[8,33,35,46] It should be mentioned here that this mobility value is an underestimated channel mobility as it includes the contact resistances at the two electrodes. Excluding such effects by the four-probe measurement method might increase the field-effect mobility.[47] However, sticking to our main aim to focus on the hysteresis at large positive gate bias, where the effect of contact resistances on the hysteresis would be not significant anymore,[28,32] all the measurements are carried out using two probe geometry to get effective mobility. The threshold voltage ($V_{TH}$) is found to be 18.85 V and is obtained from the linear curve fitting of drain current values by using the expression, $I_{ds} = \frac{W}{L}\mu_{FE}C_{ox}V_{ds}(V_{bg} - V_{TH})$.[26] In literature, it is seen that the presence of the grain boundary influences the carrier transport of the FETs and hence the mobility to a great extent due to bandgap broadening, and strain effects.[18,48] As we are getting good mobility, we expect our MoS$_2$ sample is free from such grain boundaries and the obtained mobility is only subjected to substrate effects.

The room temperature transfer characteristics for different drain-source voltages with a forward sweep of back-gate voltage in both linear (right axis) and semi-logarithmic (left axis) scale are shown in Fig. 3 (d) . The observed OFF currents are in the order of 10$^{-13}$A while the ON current for $V_{ds}$ = 100mV saturates at ~$10^{-6}$A, illustrating a magnitude of ~$10^7$ order current ON/OFF ratio for our device. Our results establish an excellent agreement with previously reported $I_{on}/I_{off}$ ratios values for back gated monolayer $MoS_2$ transistors.[8,46] Similar measurement for transfer characteristics is also carried out in ambient pressure (Air) condition with mobility and $V_{TH}$ extraction as shown in Fig. S1. A high degradation in the saturation drain current value (~10 times decrease in magnitude) is noticed with a decrease in mobility of 3.86 cm²V⁻¹s⁻¹. The current ON/OFF ratio is reduced to ~$10^4$ leaving a minor change in threshold voltage ($V_{TH}$ = 16.86V). This is expected due to the adsorption of water/$O_2$ molecules on the surface of the $MoS_2$ channel acting as charge trapping and scattering centers under the application of high positive gate voltage.[46]

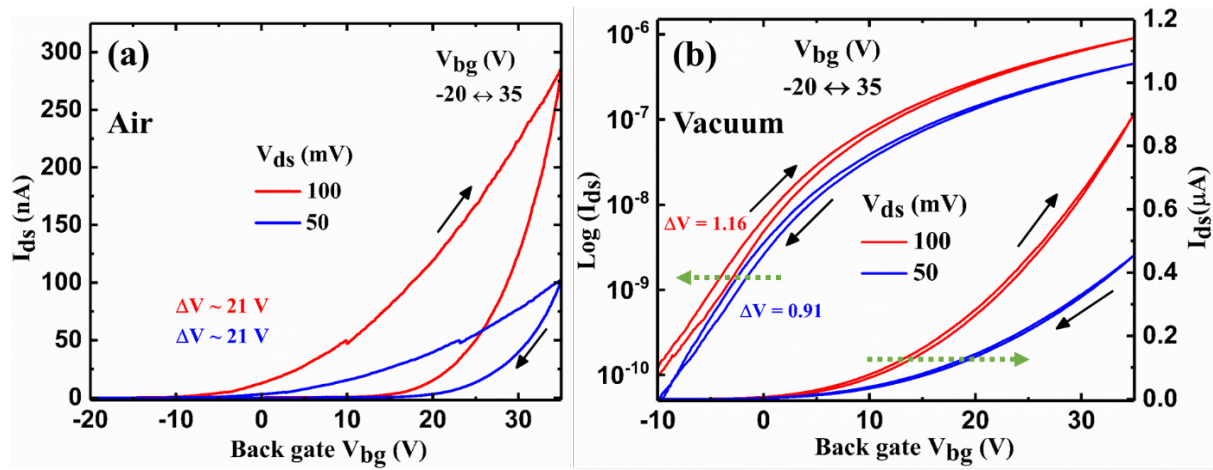

**Figure 4.** *Hysteresis loops in the transfer characteristics of $MoS_2$ FET for gate voltage sweep range -20V to 35V in **(a)** air **(b)** high-vacuum environment. The curves are shown in both linear (right axis) and semi-logarithmic scale (left axis) for two different $V_{ds}$ = 100 mV (red), 50 mV (blue). For Air condition, only linear scale is shown due to broad hysteresis. $\Delta V$ represents the maximum hysteresis width during forward and backward sweeps. Black arrows indicate the directions of drain currents with back gate sweeping voltages.*

To explore the intrinsic/extrinsic contributions to the hysteretic behavior in transfer characteristics of the MoS$_2$ FET, the DC gate voltage is further applied in a dual sweep mode. The device is first exposed to ambient pressure to observe the hysteretic nature under gate bias stress. The sweep range is varied from negative gate voltage (-20V) to high positive gate voltage (35V) and again sweep back to the negative gate voltage (-20V) in a cyclic manner as shown in Fig. 4 (a). As in literature, it is observed that voltage sweep rates can significantly influence the hysteresis of the device.[26] To avoid such additional effects a fixed sweep rate of 0.1V/s is maintained for each measurement. The observed hysteresis is measured by the maximum voltage shift in the transfer characteristics between forward and backward sweeps known as hysteresis width ($\Delta V$). The dual-sweep is recorded at two fixed drain-source voltages ($V_{ds}$) i.e. 100 mV and 50 mV to examine any possible variations on the hysteresis width. The sweeping directions induce a broad clockwise hysteresis width of $\Delta V$ ~21 V, as shown in Fig. 4 (a). This behavior is quite similar to the previous reports, which dominantly attributes to a large amount of charge trapping at the MoS$_2$ surface due to the adsorption/desorption of ambient gases and water molecules.[25,27] At the beginning of the forward sweep, when a particular negative bias is applied to the gate terminal, an initial desorption process of the trapped molecules takes place, releasing additional charge carriers to the conduction band of the MoS$_2$ channel, which causes a left shift of the transfer characteristics. As we reach a high positive gate bias, the adsorption of ambient molecules is increased, which captures more electrons filling the trapping sites. As a consequence, during the backward sweep starting from the high positive gate bias, the channel of MoS$_2$ experiences fewer number of charge carriers due to trapped charges causing degradation in drain current values, which basically leads to a right shift of the transfer characteristics. This total offset in transfer curves, also known as hysteresis width, is mainly related to two plausible factors at room temperature, i.e. (i) transfer of electrons between MoS$_2$ channel and external adsorbates,[25] (ii) oxide traps at MoS$_2$/SiO$_2$ interface,[31] which is regarded as an intrinsic factor for trapping of charges. However, to study this intrinsic trapping mechanism for better control of $MoS_2$ and other 2D material based upcoming devices, such a large amount of hysteresis caused by extrinsic molecules/gases is certainly avoidable. To rule out these external factors, we proceed with our hysteresis study in a high-vacuum (~$10^{-5}$ mbar) environment keeping the same above gate bias sweeping conditions as shown in Fig. 4 (b). Interestingly, a substantial decrease in the hysteresis width of ~95% is noticed upon switching the environment from ambient conditions to high-vacuum. The considerable reduction of $\Delta V$ in high-vacuum, not only provides the testimony of complete elimination of adsorbate effects but also demonstrates a clean, robust channel surface for the realization of intrinsic transistor parameters. The residual hysteresis is now merely attributed to the oxide trapping at the $MoS_2/SiO_2$ interface. These charge traps essentially come from unpreventable dangling bonds at the $SiO_2$ surface.[31] Another plausible factor is that presence of trapped H-bonded water molecules between $MoS_2$ and $SiO_2$ interface may additionally contribute to the charge traps as described in earlier reports.[25,29,31] However, in our case, we don't expect trapped water molecules between the $MoS_2/SiO_2$ interface as the samples are prepared at high-temperature CVD method.

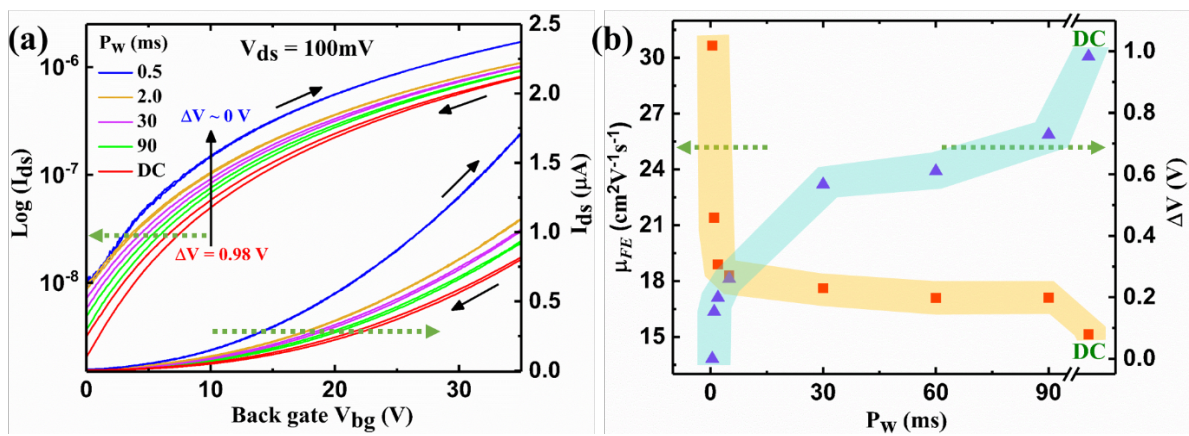

**Figure 5.** *(a) Pulsed I~V characterization with short-pulse amplitude dual-sweeping technique for base voltage fixed at 0V. Pulse amplitude is swept monotonically at a step of 0.1V to the highest amplitude of 35V at a fixed DC bias of $V_{ds} = 100$ mV. The transfer curves are obtained*

*for gate voltage pulse widths, $P_w$ = 90, 30, 2, 0.5 ms, and corresponding drain currents are shown in both semi-logarithmic (left axis) and linear (right axis) scales. A DC transfer characteristic is shown with the same sweeping rate and dual-sweeping range for the comparison. The vertical black arrow indicates a decrease in hysteresis widths (ΔV) from DC to pulse currents with decreasing pulse widths. **(b)** Variation of mobility (left axis) and hysteresis width (right axis) with different pulse widths, $P_w$, and comparison with DC mobility.*

In order to obtain the intrinsic transfer characteristics of the MoS$_2$ FET, a further study on the remnant hysteresis induced by oxide-based traps is certainly required. As per the conventional DC measurement technique is concerned, the trapping/de-trapping process at the MoS$_2$/SiO$_2$ interface is inevitable for the complete removal of hysteresis in the transfer characteristics due to the more extended measurement periods.[49,50] However, in this situation, the pulsed I~V technique is demanded due to its shorter measurement time, which might truncate the effect of trapping hence minimizing the hysteresis further.[33] Recently, numerous approaches have been addressed to quantify the charge trapping densities in bulk gate oxide as well as at the interface of MoS$_2$ and gate-dielectric by using pulsed I~V methodology.[51,52] A high-temperature hysteresis study by pulsed measurements with the formation of memory step due to slow relaxation of trapped charges indicates elevated temperature applications of MoS$_2$ thin-film devices.[53] Furthermore, controllable hysteresis with different gas pressure conditions corroborates MoS$_2$ as a versatile candidate for broad range memory-based applications.[54] Therefore, a room temperature pulse I~V measurements are carried out to inspect the intrinsic charge trapping mechanism causing hysteresis. The pulsed gate voltages with different pulse widths ($P_w$) are deployed to study the change in hysteresis at a fixed drain-source DC bias ($V_{ds}$ = 100mV). Figure 5 (a) shows the dual sweep transfer characteristics by using the pulse gate voltages with pulse widths varying from 0.5 ms to 90 ms. However, in the pulse gate voltage amplitude sweep technique, the corresponding drain currents are recorded using a spot mean measurement method in the plateau region of each pulse, as shown in S2. In order to circumvent the hysteresis in DC measurement and obtain intrinsic trap-free drain currents, the voltage pulse widths are optimized in the millisecond (ms) time-domain range. The base voltage is set to 0V in each measurement and the pulse amplitude is monotonically increased up to 35V at the step of 0.1V followed by a reverse sweep, in the same manner, generating a complete set of $I_{ds} \sim V_{bg}$ curve. The DC hysteresis is also shown to collate with pulsed hysteresis measurements having the highest hysteresis width of $\Delta V = 0.98$V. To obtain the hysteresis as a function of pulse width only, the rise and fall time of the pulses are maintained as short as possible at 1μs. From Fig. 5 (a) it can be seen that the hysteresis width ($\Delta V$) is kept on decreasing as we decrease the gate voltage pulse width from 90 ms to 2 ms and at a particular pulse width ($P_w = 0.5$ ms) it practically vanishes. At the pulse width of 0.5 ms, the backward sweep drain currents overlap with the forward sweep drain current values eliminating hysteresis at the same time providing near intrinsic transfer characteristics for the MoS$_2$ FET. On comparing to the previous report,[33] where a maximum of 89% reduction of hysteresis is observed by pulsed I~V technique with similar device configuration. However, a ~100% elimination of hysteresis in our case is observed for the first time in the case of NaCl aided CVD-grown monolayer MoS$_2$ FET. The right Y-axis of Fig. 5 (b) illustrates the variation of hysteresis widths of the transfer curves obtained by pulse I~V measurements with different pulse widths and DC measurements. In the next section, a detailed discussion on the trapping mechanism with time constants will be in order. The population of trapped charges ($N_t$ in cm$^{-2}$) can be calculated by using the equation $N_t = \Delta V \times \frac{C_{ox}}{q}$, where, $C_{ox}$ is the oxide capacitance in F/m$^2$, q being the elementary charge in C.[29] In the case of DC I~V, the hysteresis width of $\Delta V = 0.98$V corresponds to ~7$\times 10^{10}$ cm$^{-2}$ number of trapped charges present at the interface of MoS$_2$ and gate dielectric. However, in the case of pulsed I~V, the number of trapped charges are reduced to 5.2$\times 10^{10}$, 4 $\times 10^{10}$, and $1.3 \times 10^{10}$ cm$^{-2}$ for $P_w$ = 90, 30, and 2 ms, respectively. For $P_w$ = 0.5 ms a vanishing hysteresis suggests the absence of trapped of charges during forward and backward sweeps. The physical apprehension of hysteresis-free drain currents interprets higher mobility values due to the trap-free flow of charge carriers through the MoS$_2$ channel as compared to the stressed DC measurements. The left Y-axis of Fig. 5 (b) describes the pulse width dependent field-effect carrier mobility values and a fair comparison with DC mobility. The DC mobility is calculated to be 15.13 cm$^2$V$^{-1}$s$^{-1}$, whereas the pulsed mobility with pulsed width, $P_w$ = 90 ms is extracted to be 17.11 cm$^2$V$^{-1}$s$^{-1}$, and is keep on increasing until we get a trap-free near intrinsic mobility of around 30.65 cm$^2$V$^{-1}$s$^{-1}$ at a pulse width, $P_w$ = 0.5 ms. All the mobility values are calculated using forward sweep transfer curves. It is worth noticing that the mobility extracted from the hysteresis-free drain currents (corresponds to $P_w$ = 0.5ms) is increased by ~100% than the DC mobility value elucidating considerable current enhancement on the application of pulsed gate bias sweep as compared to the previous report.[33]

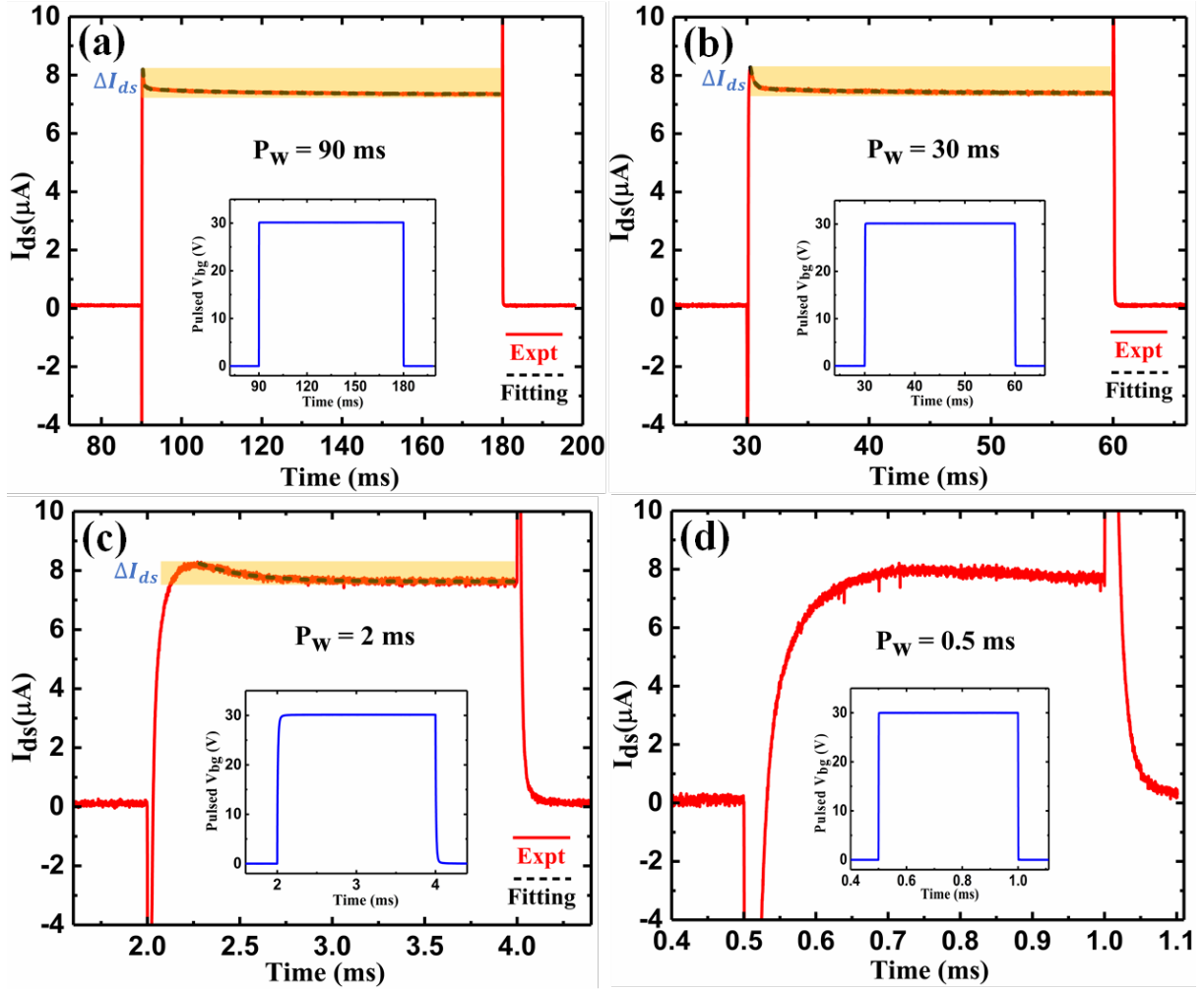

**Figure 6.** *Transient drain current characteristics (red curve) by using single pulse waveform measurement for different pulse widths, $P_w$ = (a) 90 ms (b) 30 ms (c) 2 ms (d) 0.5 ms. Insets show the pulse gate voltage (blue curve) from 0V to 30V as a function of time. The drain current reduction ($\Delta I_{ds}$) in each case is highlighted in yellow color. The fittings to the drain currents are shown in the black dotted line by using single and double exponential decay equations given in the main text.*

During a pulse I~V measurement, the hysteresis minimization is obtained by considering two possible approaches. Here, we discuss them one by one in detail. Case-1: A pulse voltage in waveform has three components i.e., rise time, pulse width, and fall time consisting of a pulse period. The time difference between the two pulses (i.e., falling edge of one pulse and the rising edge of the next pulse) is denoted as delay time, as shown in S2. Charge trapping occurs during the pulse width, whereas throughout the delay time de-trapping process is expected.[31] As it is a well-known fact that de-trapping is a much slower process than trapping.[52] However, if one keeps the delay time the same as pulse width (which is in our case), the number of trapped charges will not get enough time to recover. Now during the pulse amplitude sweep of the gate voltage, as discussed in Fig 5 (a), we apply a pulse train with monotonically increasing its amplitude. During the forward sweep, the accumulation of charges will occur at the trapping sites because, in our pulse time parameters, the pulse width and delay time are the same. This accumulation of trapped charges causes hysteresis in the transfer curve by pulse I~V measurements. Now, one way to avoid this accumulation is by giving more time for the charges to be de-trapped i.e., increasing the delay time. In the previous reports, various research groups have opted for this method to reduce the hysteresis in transfer characteristics using pulsed I~V measurements. Although using a much longer delay time minimizes the hysteresis a little bit further, however, none of the consensuses have ever been reported to completely rule out the hysteresis by this method.[29,33,52]

Case-2: In this context, we propose another approach which not only eliminates the hysteresis completely but also provides a faster measurement method for getting intrinsic transistor properties. As shown in Fig. 5 (a), by decreasing the pulse width of the gate voltage pulse train and simultaneously measuring the drain current values produce transfer curves with reduced hysteresis and as the pulse width is sufficiently low i.e. at $P_w = 0.5$ ms the hysteresis is eliminated giving rise to high ON-current values as compared to other transfer curves. To have a clear understanding, we switch to a waveform transient drain current analysis by employing a single voltage pulse to the gate terminal, as shown in Fig. 6. The drain-source voltage is fixed at 1V for each measurement. The pulse width is varied from 90 to 0.5 ms to inspect the minimum voltage stress time required to induce charge trapping significantly at the $MoS_2/SiO_2$ interface. The humps in the current values at the pulse transitions as seen from the figure, stem from capacitive charging effects from the higher cable capacitance.[55] The block diagram illustrating this fundamental effect is shown in S3. At the rising edge of the pulse from 0V to 30V, the drain current increases suddenly to ~8 μA due to the accumulation of charge carriers into the $MoS_2$ channel. At the beginning of the plateau region of the pulse voltage, the drain current drops quickly indicating an initial fast charge trapping followed by a slower, continuous decrease in current values with the voltage pulse maintained at 30V. The reduction of drain current ($\Delta I_{ds}$), shown in the highlighted region (yellow), basically attributes to some of the channel electrons is getting captured at the trapping sites due to the voltage stress during the pulse width. As seen in Fig. 6, this degradation of the drain current ($\Delta I_{ds}$) is more in case of wider pulse width ($P_w$ = 90ms), stipulating more trapping of charges, which gradually decreases with the decrease in pulse widths. However, the transient curve associated with the pulse voltage having the pulse width, $P_w$ = 0.5 ms experiences no degradation in the current values elucidating inadequate response time for electrons to be trapped. These results explain the reason behind the shifting of the ON-current values for different pulse width conditions corresponds to the spot-mean degraded current values except for $P_w$ = 0.5 ms, where a high, hysteresis-free drain current values are recorded. This is the required fastest pulse, which might provide actual drain current values, free from any significant charge trapping effects and hence hysteresis, representing intrinsic transistor behavior.

The fleeting drain current values at the plateau region of the pulse voltage are fitted with two trap models for $P_w$ = 90 and 60 ms whereas for $P_w$ = 2 ms, the drain current values are best fitted with one trap model as given by the following equations.

Two trap model: $I = I_0 + A(e^{-(t-t_0)/\tau_1}) + B(e^{-(t-t_0)/\tau_2})$            (1)

One trap model: $I = I_0 + A(e^{-(t-t_0)/\tau_1})$            (2)

Where, $\tau_1$ and $\tau_2$ are the time constants for fast and slow trapping processes, respectively. The parameters A and B are initial amplitudes (currents) for the two trapping contributions, $I_0$ and $t_0$ are adjustable offsets in current and time frames, respectively. The values of these fitting parameters and constants for different pulse widths are shown in the table.

| $P_w$ (ms) | A (Amp.) | B (Amp.) | $\tau_1$ (s) | $\tau_2$ (s) |
|---|---|---|---|---|
| 90 | $6.6 \times 10^{-7}$ | $1.8 \times 10^{-7}$ | $3.2 \times 10^{-4}$ | $2.6 \times 10^{-2}$ |
| 30 | $7.1 \times 10^{-7}$ | $1.6 \times 10^{-7}$ | $2.8 \times 10^{-4}$ | $1.0 \times 10^{-2}$ |
| 2 | $6.0 \times 10^{-7}$ | - | $2.6 \times 10^{-4}$ | - |

The two aforementioned models appear to fit suitably the entire region of transient currents when the transistor is turned ON by the pulse voltage. The trapping time constants indicate the occurring of multiple trapping processes for different pulse width conditions. The faster trapping process may attribute to the tunneling of electrons at the $MoS_2/SiO_2$ interface sub-gap states, whereas the relatively longer trapping process may arise due to the jumping of channel electrons into deeper defect states of $SiO_2$.[51,54] However, with decreasing the pulse width to 2 ms, the longer trapping process is eliminated due to insufficient time for the tunneling of electrons into deeper states. Our findings on NaCl assisted CVD grown monolayer $MoS_2$ FET not only unfold the fundamentals of charge trapping mechanism but also offer a meticulous approach towards intrinsic, hysteresis-free transistor performances by using pulse I~V and fast transient measurement techniques for practical device applications.

## Conclusions

In summary, a progressive study on the time-domain drain current responses is demonstrated to minimize the fast transient charge trapping effects at $MoS_2/SiO_2$ interface in a NaCl aided CVD-grown monolayer $MoS_2$ FET. The role of the Ag as a contact metal for drain/source electrode is critically analyzed by using a comprehensive DFT study to explore the underneath interfacial properties. The Schottky barrier at the Ag/MoS$_2$ interface is calculated to be 153 meV from the projected band profile of the heterostructure. The room temperature hysteresis analysis using DC and pulse I~V techniques is performed in both ambient and vacuum conditions to realize different trapping mechanisms. An increase in the field-effect mobility by 100% is perceived along with hysteresis-free, intrinsic transfer characteristics in the case of short pulse amplitude sweep measurement than the corresponding DC measurement. Finally, single-pulse transient measurements are carried out to get insight into the fast and slow trapping phenomena at the interface between MoS$_2$ and gate-dielectric. Our experimental results are helpful in achieving the fundamentals of trapping mechanisms to obtain the enhanced trap-free channel mobility value in the case of monolayer MoS$_2$. This proposed study indicates the importance of controlling the interfacial charge traps for further advancement of the device performances, not only in monolayer MoS$_2$ but also in other two-dimensional materials and its heterostructures.

## Methods

### *CVD synthesis and characterization of MoS$_2$*

Single-layer MoS$_2$ is synthesized using a single zone tube furnace (Carbolite 1200) in a 2-inch diameter quartz tube. For this specific study, MoO$_3$ (99.9995%, Alfa Aesar) and sulfur (99.9995%, Alfa Aesar) powder are taken as precursors in two separate alumina boats. A small amount of NaCl (>99%) is added to the MoO$_3$ for salt assisted CVD synthesis. SiO$_2$(300nm)/Si (1cm×1cm) substrate is sequentially cleaned with acetone, IPA, DI water in the ultrasonication method followed by treated with pressurized dry N$_2$ (99.999%) to remove moistures. The cleaned substrate was placed on the boat facedown to the MoO$_3$ powder and pushed to the center of the furnace. The boat containing sulfur is placed in an upstream region. Initially, the tube is pumped down to 0.2mbar and then purged three times with 500 sccm high pure Ar (99.999%) to remove oxygen contents in the tube. The system is heated to 750 °C at a ramp speed of 20 °C/min with Ar (80 sccm) as the carrier gas. The monolayer $MoS_2$ is synthesized at 750 °C for 3 mins and the pressure is maintained at 600 mbar. Finally, the system was cooled down quickly to ambient temperature by partially opening the furnace. Raman measurements are carried out using a micro-Raman spectrometer (T-64000, Horiba Jobin Yvon) with 514.5 nm laser line of an Ar ion laser as the excitation source and 100X objective.

### *Device fabrication and electrical measurements*

The single-layer $MoS_2$ FETs are fabricated by using photolithography (Heidelberg µPG 101) system. Firstly, the $MoS_2$ grown $SiO_2$/Si substrate is coated with a positive photoresist (ma-p-1205) by spin coater (SUSS Microtech) and then baked at 80 °C for 1 min. Under the inspection of a high-resolution microscope, the required contact patterns are exposed on the single-layer $MoS_2$ flakes with a 405 nm laser in photolithography. The exposed patterns are developed with an alkaline solution (1:4, NaOH: DI water) and the sample is mounted in a thermal evaporation chamber for the deposition of the contact metal (Ag) followed by a lift-off process. The back gate of the FETs is facilitated by the heavily doped Si substrate. Finally, The devices are mounted on the micro manipulating four-probe stage (Lakeshore cryogenic probe station) for probing the contacts, the DC and pulsed electrical measurements are recorded by Keithley 4200-A SCS parameter analyzer both in air and high-vacuum ($\sim 10^{-5}$ mbar) conditions. We limited our measurements to $V_{bg}$ (back gate voltage) $\sim 38$ V (with compliance) to keep the device safe for more extended measurements. In order to prevent photo-excitation of charge carriers, all the experiments are performed in dark conditions.

### *Computational details*

All the electronic calculations are performed within the density functional theory (DFT) by using the Quantum Espresso software package. To include the interaction between the core and the valence electrons, the ultrasoft pseudopotential (USPP) along with a plane wave basis set is considered for the calculations and the exchange-correlation interaction of the system is included by Perdew-Burke-Ernzerhof (PBE) functional. Keeping in sight on reduced computational cost, plane wave cut off energy is taken to be 80 Ry and the cut off energy for the charge density is chosen as 800 Ry as obtained during the optimization process. All the structures are optimized by the conjugate-gradient minimization scheme. The Brillouin zone integrations are carried out within an un-shifted Monkhorst-pack k points mesh of $10 \times 10 \times 1$ and $6 \times 6 \times 1$ in the case of the unit cell and supercell, respectively. During the structural relaxation, the ion dynamics are regulated under BFGS algorithms. The lattice constant for the relaxed monolayer unit cell is optimized to be 3.188 Å. A sufficient vacuum in the z-direction is provided in all the calculated structures to avoid the interaction between two layers. The self-consistent calculations are performed until a total energy threshold convergence of $10^{-6}$ Ry and the force on the atoms are converged up to a difference of $10^{-3}$ Ry/Å. The heterostructure of Ag (111) and $MoS_2$ is constructed with $3 \times 3 \times 1$ supercell ensuring a small lattice mismatch and an equilibrium separation is fixed at 2.5 Å as per the energy minimization. The orbital projected band structure and DOS calculations are performed with the abovementioned parameters in case of unit cell and supercell.

## Acknowledgment

We acknowledge Shikha Varma for extending the Raman facility and S. Choudhury for assistance during the Raman experiment. SKG would like to thank the Science and Engineering Research Board (SERB), India for financial support (Grant no.: YSS/2015/001269). R.A. thanks to Swedish Research Council (VR-2016-06014) for financial support. SNIC, HPC2N, at Sweden and SAMKHYA, at IOP, Bhubaneswar are acknowledged for providing the computing facilities.

temperature performance of MoS$_2$ thin-film transistors: Direct current and pulse current-voltage characteristics. *J. Appl. Phys.* **117**, 064301 (2015).